\documentclass[reprint,amsmath,amssymb,aps,superscriptaddress]{revtex4-2}
\usepackage{graphicx}
\usepackage{xfrac}
\usepackage{xcolor}
\usepackage{bm}
\usepackage{hyperref}

\def\CRA{CeRh$_2$As$_2$}

\newcommand{\orange}[1]{\textcolor{black}{#1}}
\newcommand{\red}[1]{\textcolor{black}{#1}}

\begin{document}
\preprint{APS/123-QED}

\title{Quasi-Two-Dimensional Antiferromagnetic Spin Fluctuations in the Spin-Triplet Superconductor Candidate \CRA}

\author{Tong Chen}
\email{tchen115@jhu.edu}
\affiliation{Institute for Quantum Matter and Department of Physics and Astronomy, Johns Hopkins University, Baltimore, Maryland 21218, USA}

\author{Hasan Siddiquee}
\affiliation{Department of Physics, Washington University in St. Louis, St. Louis, Missouri 63130, USA}

\author{Qiaozhi Xu}
\affiliation{Department of Physics, Washington University in St. Louis, St. Louis, Missouri 63130, USA}

\author{Zack Rehfuss}
\affiliation{Department of Physics, Washington University in St. Louis, St. Louis, Missouri 63130, USA}

\author{Shiyuan Gao}
\affiliation{Institute for Quantum Matter and Department of Physics and Astronomy, Johns Hopkins University, Baltimore, Maryland 21218, USA}

\author{Chris Lygouras}
\affiliation{Institute for Quantum Matter and Department of Physics and Astronomy, Johns Hopkins University, Baltimore, Maryland 21218, USA}

\author{Jack Drouin}
\affiliation{Institute for Quantum Matter and Department of Physics and Astronomy, Johns Hopkins University, Baltimore, Maryland 21218, USA}

\author{Vincent Morano}
\affiliation{Institute for Quantum Matter and Department of Physics and Astronomy, Johns Hopkins University, Baltimore, Maryland 21218, USA}

\author{Keenan E. Avers}
\affiliation{Maryland Quantum Materials Center and Department of Physics, University of Maryland, College Park, Maryland 20742, USA}

\author{Christopher J. Schmitt}
\affiliation{Neutron Scattering Division, Oak Ridge National Laboratory, Oak Ridge, Tennessee 37831, USA}

\author{Andrey Podlesnyak}
\affiliation{Neutron Scattering Division, Oak Ridge National Laboratory, Oak Ridge, Tennessee 37831, USA}

\author{\red{Johnpierre Paglione}}
\affiliation{Maryland Quantum Materials Center and Department of Physics, University of Maryland, College Park, Maryland 20742, USA}

\author{Sheng Ran}
\email{rans@wustl.edu}
\affiliation{Department of Physics, Washington University in St. Louis, St. Louis, Missouri 63130, USA}

\author{Yu Song}
\email{yusong\_phys@zju.edu.cn}
\affiliation{Center for Correlated Matter and School of Physics, Zhejiang University, Hangzhou 310058, China}

\author{Collin Broholm}
\email{broholm@jhu.edu}
\affiliation{Institute for Quantum Matter and Department of Physics and Astronomy, Johns Hopkins University, Baltimore, Maryland 21218, USA}
\affiliation{Department of Materials Science and Engineering,
The\ Johns\ Hopkins\ University, Baltimore, Maryland\ 21218, USA}

\date{\today}


\begin{abstract}
The tetragonal heavy-fermion superconductor \CRA{} ($T_{\rm c}=0.3$~K) exhibits an exceptionally high critical field of  14~T for ${\bf B} \parallel {\bf c}$. It undergoes a field-driven first-order phase transition between superconducting states, potentially transitioning from spin-singlet to spin-triplet superconductivity. To \orange{further understand these superconducting states and the role of magnetism,} we probe spin fluctuations in \CRA{} using neutron scattering. We find dynamic $(\pi,\pi)$ antiferromagnetic (AFM) spin correlations with an anisotropic quasi-two-dimensional correlation volume. Our data place an upper limit of 0.31~$\mu_{\rm B}$ on the staggered magnetization of corresponding N\'{e}el orders at $T=0.08$~K. Density functional theory calculations, treating Ce $4f$ electrons as core states, show that the AFM wave vector connects significant areas of the Fermi surface. Our findings indicate that the dominant excitations in \CRA{} for $\hbar\omega< 1.2$~meV are magnetic and suggest that superconductivity in \CRA{} is mediated by AFM spin fluctuations associated with a proximate quantum critical point.
\end{abstract}

\maketitle


While phonon-mediated superconductivity is typically incompatible with magnetism, antiferromagnetic (AFM) spin fluctuations can promote unconventional superconductivity \cite{moriya2000spin, scalapino2012common} in various systems, including cuprates \cite{rossat1991neutron, mook1993polarized, fong1995phonon, tranquada2014superconductivity}, Fe-pnictides \cite{christianson2008unconventional, inosov2010normal, dai2015antiferromagnetic} and chalcogenides \cite{wang2016strong, chen2019anisotropic}, and heavy-fermion metals \cite{smidman2023colloquium}. Experimental evidence for magnetically driven superconductivity includes (i) spin resonance modes in the superconducting (SC) state \cite{stock2008spin, stockert2011magnetically, duan2021resonance}, and (ii) paramagnetic excitations in the normal state with energies well beyond the SC gap \cite{regnault1988neutron, arndt2011spin, wang2016magnetic, song2020nature, song2021high}. Despite extensive efforts to understand the spin dynamics of spin-singlet unconventional superconductors, research on spin-triplet superconductors has been constrained by the scarcity of model systems \cite{duan2021resonance, carr2016diverse}. 

\begin{figure}[t]
    \centering
    \includegraphics[width=1\columnwidth]{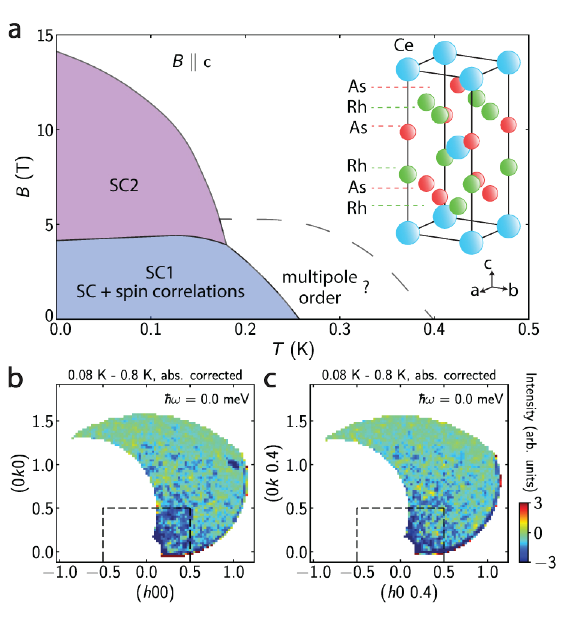}
    \caption{
    (a) Schematic field-temperature phase diagram of \CRA{} for ${\bf B}\parallel {\bf c}$ based on Refs.~\cite{khim2021field, landaeta2022field, mishra2022anisotropic, onishi2022low, hafner2022possible, semeniuk2023decoupling, chajewski2024discovery}. The inset shows the crystal structure of \CRA.
    $\bf Q$-dependence of the difference in elastic neutron scattering between $T=0.08$~K and $T=0.8$~K for (b) $|l|<0.15$ and (c) $0.35 < |l| < 0.45$.
    Dashed lines indicate the Brillouin zone. 
    } 
    \label{fig:fig1}
\end{figure}

\CRA{} is a heavy-fermion superconductor with $T_{\rm c} = 0.3$~K \cite{khim2021field}. It adopts the tetragonal CaBe$_2$Ge$_2$-type structure (space group No. 129 P4/nmm), consisting of two-dimensional (2D) Ce layers stacked with As-Rh-As and Rh-As-Rh blocks along the ${\bf c}$ axis [Fig.~\ref{fig:fig1}(a) inset]. While the overall crystal structure is centrosymmetric, Ce layers are positioned between distinct blocks, leading to local inversion symmetry breaking and alternating Rashba spin-orbit coupling (SOC) \cite{bihlmayer2022rashba, fischer2023superconductivity}. For ${\bf B} \parallel {\bf c}$, \CRA{} exhibits pronounced anomalies in alternating current susceptibility, magnetization, and magnetostriction \cite{khim2021field, landaeta2022field, mishra2022anisotropic, onishi2022low, hafner2022possible, semeniuk2023decoupling, chajewski2024discovery}, indicating a first-order field-driven transition between different SC states [Fig.~\ref{fig:fig1}(a)]. Similar phase transitions within the SC state were observed in  UPt$_3$ \cite{aeppli1989magnetic} and CeCoIn$_5$ \cite{yoshida2012pair} and attributed to the interplay between superconductivity and competing magnetic order parameters.  In \CRA{}, the presence of alternating Rashba SOC suggests the possibility of a transition from spin-singlet to spin-triplet superconductivity \cite{schertenleib2021unusual, skurativska2021spin, nogaki2021topological, mockli2021superconductivity, cavanagh2022nonsymmorphic, nogaki2022even, cavanagh2023pair, yanase2022topological, suh2023superconductivity, szabo2024effects}\orange{, although other scenarios may also apply \cite{machida2022violation, hazra2023triplet}.} 

\orange{To further understand the superconducting phases of \CRA{}, we} use neutron scattering to probe the low-temperature ($T$) zero-field magnetism of \CRA. 
Our elastic scattering data reveal no significant differences between 0.08~K and 0.8~K, indicating the absence or weakness of magnetic order, and place an upper limit of 0.31~$\mu_{\rm B}$ on the staggered magnetization of corresponding N\'{e}el orders at temperatures down to 0.08~K. 
Our inelastic magnetic neutron scattering data, however, reveal a gapless spectrum of quasi-2D AFM spin fluctuations at the $(\pi,\pi)$ wave vector $\textbf{Q}_{\rm AFM}$ extending up to 1.2~meV. In combination with density functional theory (DFT) calculations  \cite{ptok2021electronic, nogaki2021topological, cavanagh2022nonsymmorphic, hafner2022possible, wu2024fermi,chen2024coexistence, chen2024exploring, ishizuka2024correlation} and angle-resolved photoemission spectroscopy (ARPES) measurements \cite{wu2024fermi, chen2024coexistence, chen2024exploring} that reveal Fermi-surface nesting at $\textbf{Q}_{\rm AFM}$, our findings suggest proximity to Fermi-surface-driven magnetic criticality in \CRA{} \cite{khim2021field, hafner2022possible}, with the associated spin fluctuations playing a crucial role in the superconductivity.

\CRA{} single crystals were synthesized via a flux method \cite{khim2021field, siddiquee2023pressure}. The sharp resistive transition of a representative sample at $T_{\rm c} = 0.3$~K is displayed in the Supplemental Material \cite{SupplementalMaterial}. For neutron scattering experiments in the $(hk0)$ plane, $\sim 120$ crystals (0.9~g) were coaligned with a full width at half maximum (FWHM) mosaic width of 2.8~degrees for rotation around the ${\bf c}$ axis. As detailed in Supplemental Material \cite{SupplementalMaterial}, precautions were taken to ensure good thermal contact between the dilution refrigerator and the sample.
Neutron experiments were conducted at Oak Ridge National Laboratory using the CNCS Spectrometer \cite{ehlers2011new} with a fixed incident energy $E_i$ of 3.32~meV. Operating in the high flux mode with the disk chopper rotating at 180~Hz, this setup resulted in an FWHM elastic energy resolution of 0.17~meV.

Normalization of the measured scattering cross sections was achieved by comparing the count rate with the $\bf Q$-integrated count rate for the (110) nuclear Bragg peak. As detailed in the Supplemental Material \cite{SupplementalMaterial}, a scattering angle and energy transfer dependent factor was applied to background subtracted data to correct for the significant absorption cross section of \CRA{}. When background subtraction is not feasible, uncorrected intensity data are presented in arbitrary units. In the following, momentum transfer $\textbf{Q}$ is expressed as $\textbf{Q}=h{\bf a^*}+k{\bf b^*}+l{\bf c^*}$, where $(hkl)$ denotes Miller indices. A pair of indices $(hk)$ specify the in-plane component of wave vector $\bf Q_\parallel$. Here,  ${a^*}={b^*}=2\pi/a$ and ${c^*}=2\pi/c$, with $a=b=4.28$~\AA{} and $c=9.86$~\AA{} at room temperature \cite{khim2021field}.

The difference in elastic scattering between data acquired in the SC state at 0.08~K and in the normal state at 0.8~K is shown in Figs.~\ref{fig:fig1}(b) and~\ref{fig:fig1}(c) for $l=0$ and $l=\pm0.4$, respectively. 
No diffraction peaks are apparent in either figure. These data place upper limits on magnetic ordering at specific magnetic wave vectors (see Table~\ref{table} and the Supplemental Materia \cite{SupplementalMaterial}).

\begin{table}
\begin{center}
\caption{Upper limits on the magnitudes of staggered magnetization in \CRA{} obtained from the difference between elastic neutron scattering at $T=0.08$~K and $T=0.8$~K shown in Fig.~\ref{fig:fig1}(b)-\ref{fig:fig1}(c). In-plane oriented magnetic order is harder to detect due to the domain averaged polarization factor}
\label{table}
\begin{tabular}{|c|c|c|} 
\hline
${\bf Q}_m$ &  $m_{z,{\rm max}}$ ($\mu_{\rm B}$) & $m_{\parallel,{\rm max}}$ ($\mu_{\rm B}$)\\
\hline
($\frac{1}{2}\frac{1}{2}0$) &  0.10 & 0.14\\
($\frac{1}{2}\frac{1}{2}\frac{1}{2}$) &  0.14 & 0.20\\
(010) &  0.14 & 0.20\\
(01$\frac{1}{2}$) &  0.22 & 0.31\\
quasi-2D ($\frac{1}{2}\frac{1}{2}$) & 0.14 & 0.20\\
 \hline
\end{tabular}
\end{center}
\end{table}

Nuclear quadrupole resonance (NQR) \cite{kibune2022observation} and nuclear magnetic resonance (NMR) \cite{ogata2023parity} experiments provide evidence for an increase of the NQR and NMR linewidths in the low-field SC1 state for the As sites of As-Rh-As blocks, which has distinct Ce neighbors displaced along ${\bf c}$, but not for the As sites of Rh-As-Rh blocks, where Ce neighbors come in groups of four whose fields cancel out in the AFM states shown in Table~\ref{table}. A possible explanation is weak long- or short-range AFM ordering that is detectable with local probes such as NMR and NQR \cite{kitagawa2022two,ogata2023parity}, but sufficiently weak and/or broad in $\bf Q$ to evade detection through neutron diffraction (see Table~\ref{table}). 


\begin{figure}[t]
    \centering
    \includegraphics[width=1\columnwidth]{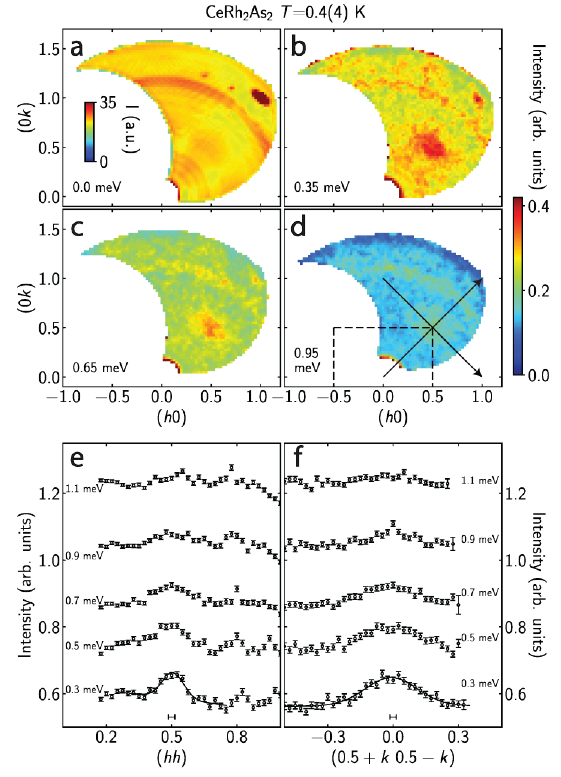}
    \caption{
    (a)-(d) $\bf Q_\parallel$-dependence of neutron scattering from \CRA{} at selected $\hbar\omega$ for $T=0.4(4)$~K and $|Q_z|<0.5c^*$. The energy windows are $0.4$~meV in (a) and $0.3$~meV in (b)-(d).
    Arrows indicate the longitudinal $(hh)$ and transverse $(k\bar{k})$ directions.
    (e),(f) Constant energy cuts along $(hh)$ and $(k\bar{k})$ through $\textbf{Q}_{\rm AFM}=(\frac{1}{2}\frac{1}{2})$ for selected $\hbar\omega$ with an energy window of $0.2$ meV. The averaging windows are $\Delta^2{\bf Q}=\pm (0.15,-0.15,0)\pm (0,0,0.5)$ in (e) and $\Delta^2{\bf Q}=\pm (0.1,0.1,0)\pm (0,0,0.5)$ in (f). 
    Solid lines for $\hbar\omega =0.3$~meV  are Gaussian fits. The horizontal bars represent the FWHM resolution. 
    } \label{fig:fig2}
\end{figure}

\begin{figure}[t]
    \centering
    \includegraphics[width=1\columnwidth]{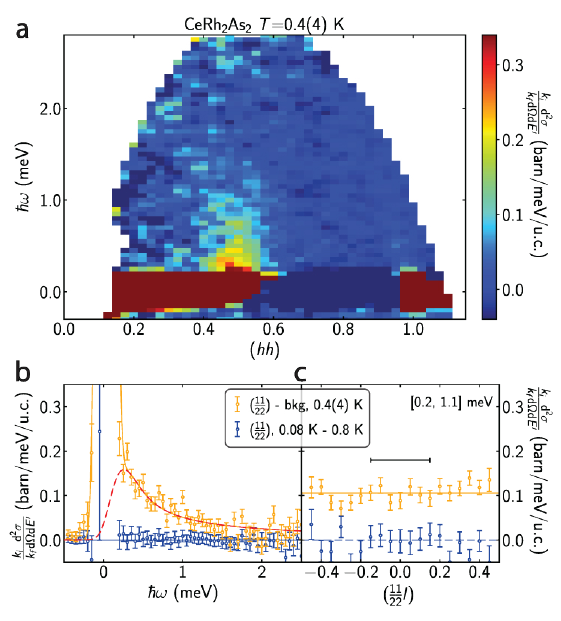}
    \caption{ 
    (a) Background subtracted inelastic neutron scattering versus ${\bf Q}=(hh)$ and $\hbar\omega$ for \CRA{} at $T=0.4(4)$~K. The signal averaging window is $\Delta^2{\bf Q}=(\pm (0.15,-0.15),\pm 0.5)$. The $|\bf Q|$ and $\hbar\omega$ dependent background was determined by averaging all data outside the signal volume. 
    (b,c) Background subtracted data at $T=0.4(4)$~K in orange and the difference between $T=0.08$~K and $T=0.8$~K data in blue. (b) shows an $\hbar \omega$ scan at ${\bf Q_{\rm AFM}}=(\frac{1}{2}\frac{1}{2})$ averaging over $\Delta^3{\bf Q}=(\pm 0.1,\pm 0.1, \pm 0.5)$. The red line shows a fit to the data described in the text. (c) shows a scan along ${\bf Q}=(\frac{1}{2}\frac{1}{2}l)$ averaging over $\hbar\omega \in [0.2,1.1]$ meV and $\Delta^2{\bf Q}=\pm (0.1,0.1)\pm (0.15,-0.15)$. 
    The horizontal bar represents the resolution width.
    } 
    \label{fig:fig3}
\end{figure}

Though we do not detect elastic magnetic diffraction, dynamic AFM spin correlations are apparent in the inelastic scattering. Figures~\ref{fig:fig2}(a)-(d) show the $\bf Q$-dependence within the $(hk)$ plane of the scattering intensity for various fixed $\hbar\omega$. 
\orange{As AFM fluctuations are observed in both paramagnetic and SC states with no discernible differences within the $\hbar\omega$-range covered, measurements taken at 0.08~K and 0.8~K are averaged and denoted as 0.4(4)~K to improve statistical accuracy.}
The elastic data display the (110) nuclear Bragg peak from \CRA{}, which we use for normalization and alignment. Other $T$-independent features in the elastic scattering are associated with the sample holder (powder ring and low $|{\bf Q}|$ diffuse scattering) and two small misaligned \CRA{} crystallites. 
At $\hbar\omega=0.35$ meV, a diffuse peak is observed near the  AFM wave vector $\textbf{Q}_{\rm AFM}=(\frac{1}{2}\frac{1}{2})$  [Fig.~\ref{fig:fig2}(b)]. While the peak position persists, the intensity decreases with increasing $\hbar\omega$.

Figures~\ref{fig:fig2}(e) and ~\ref{fig:fig2}(f) show constant-$\hbar\omega$ cuts through these data along the longitudinal $(hh)$ and transverse $(k\bar{k})$ directions. These cuts show that the low-energy magnetic scattering is centered at $\textbf{Q}_{\rm AFM}$ consistent with dynamic short-ranged AFM spin correlations as in many Ce-based heavy fermion systems including SC CeCoIn$_5$ \cite{stock2008spin}. 
Fitting the cuts to Gaussians, we infer dynamic correlation lengths (defined as 2/FWHM) at 0.3~meV along ($hh$) and ($k\bar{k}$) of $\xi_{hh}=8.1(8)$~\AA{} and $\xi_{k\bar{k}}=3.4(3)$~\AA, respectively. Considering the fourfold rotation axis of paramagnetic \CRA{}, this anisotropy suggests a preferred polarization factor in the magnetic neutron scattering cross section. Specifically, scattering at ${\bf Q_{\rm AFM}}$ detects spin fluctuations in the plane perpendicular to ${\bf Q_{\rm AFM}}$, thus along $\bf c$ or $(1\bar{1}0)$, of which the latter component can give rise to the observed anisotropy. The data indicate an underlying nematic character to the AFM spin correlations, where the correlation length is longer perpendicular to than parallel to the fluctuating in-plane staggered magnetization. Such anisotropy can arise from symmetric anisotropic exchange interactions and was previously observed in iron-based superconductors\cite{dai2015antiferromagnetic}. Given the spin-orbital character of Ce-based magnetism, anisotropic interactions are indeed anticipated. The data specifically indicate in-plane anisotropic interactions and suggest that the fourfold rotation symmetry will be broken by the corresponding static order.

Figure~\ref{fig:fig3}(a) displays the $\bf Q-\omega$ dependent magnetic scattering cross section for ${\bf Q}\parallel (11)$.
A ridge of low-energy excitations at $\textbf{Q}_{\rm AFM}$ extending to 1.2 meV is apparent.
Figure~\ref{fig:fig3}(b) shows a broad constant-$\textbf{Q}_{\rm AFM}$ cut through these data. While data for $|\hbar\omega|<0.15~$meV are dominated by imperfect subtraction of the incoherent elastic scattering, we associate the tail of scattering for $\hbar\omega>0.15$~meV that is absent on the neutron energy gain side ($\hbar\omega<-0.15$~meV) with magnetic quantum fluctuations for $k_BT \ll \hbar \omega$.
To further characterize the scattering, we fit these data against a resolution-convoluted Lorentzian response function of the form :
\begin{equation}
\begin{aligned}
{\cal S}(\omega) = \frac{1}{1-e^{-\beta\hbar\omega}} \frac{\chi_0\Gamma\omega}{\Gamma^2+\omega^2} ,
\end{aligned}
\label{Eq:lorentz}
\end{equation}
plus a central Gaussian peak to account for the residual elastic background. As shown in Fig.~\ref{fig:fig3}(b), this model provides an excellent account of the data with a characteristic relaxation rate of $\hbar\Gamma=0.10(4)$~meV. This suggests proximity to a quantum critical point (QCP) where $\hbar\Gamma\sim k_BT$\cite{broholm1987spin, stockert2006peculiarities, stock2008spin, stockert2011magnetically}. 

From the total magnetic scattering associated with inelastic magnetic scattering below 2.5~meV, we obtain the dynamic moment of $1.6(4)\mu_{\rm B}^2$. 
The reduction in moment compared to the effective moment inferred from susceptibility ($\mu_{\rm eff}^2=6.55\mu_{\rm B}^2$ \cite{khim2021field}) is observed in other Ce-based heavy-fermion superconductors, such as CeCoIn$_5$ \cite{stock2008spin} and CeCu$_2$Si$_2$ \cite{stockert2011magnetically}. 
Magnetic excitations extend to much higher energies in these systems \cite{song2020nature, song2021high} due to crystal electric field and Kondo screening. Likewise, in \CRA{} we expect such interactions to define a low-energy reduced moment regime with residual AFM intersite interactions.   

To probe the associated spin correlations along ${\bf c}$, Fig.~\ref{fig:fig3}(c) shows the $l$ dependence of scattering for $\hbar\omega\in[0.2,1.1]~$meV at $\textbf{Q}_{\rm AFM}$. The corresponding count rate at ${\bf Q}_{\rm bkg}=(0.1, 0.7)$ where $|\textbf{Q}_{\rm bkg}|=|\textbf{Q}_{\rm AFM}|$ was used as background. The inelastic scattering at $\textbf{Q}_{\rm AFM}$ consistently exceeds that at ${\bf Q}_{\rm bkg}$ by a nearly $l$-independent amount, indicating quasi-2D spin correlations. No difference between scattering at 0.08 K and 0.8 K is observed in the accessible energy range.

\begin{figure}[t]
    \centering
    \includegraphics[width=1\columnwidth]{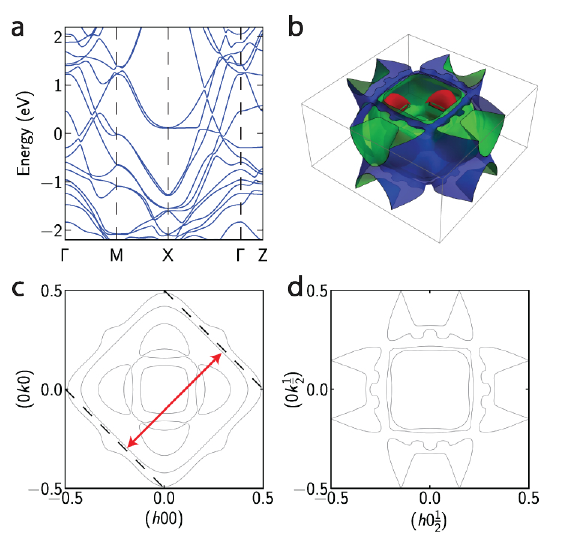}
    \caption{
    (a) Electronic band structure of \CRA{} calculated using DFT with Ce $4f$ electrons in the core. (b)-(d) The corresponding Fermi surface as a 3D image and as 2D slices through (c) the $(hk0)$ plane and (d) the $(hk\frac{1}{2})$ plane. The red arrow indicates the wave vector ${\bf Q}_{\rm AFM}=(\frac{1}{2}\frac{1}{2})$ associated with low energy inelastic magnetic neutron scattering [Fig.~3(a)], which is seen to connect significant portions of the DFT Fermi surface.
    } \label{fig:fig4}
\end{figure}

To gain insights into the origin of the AFM spin fluctuations in \CRA, we conducted band structure calculations using DFT, treating Ce 4$f$ electrons as core states. The results are shown in Fig.~\ref{fig:fig4}(a), with the corresponding Fermi surface displayed in Figs.~\ref{fig:fig4}(b)-\ref{fig:fig4}(d).
Specifically, the Fermi surface at $l=0$ [Fig.~\ref{fig:fig4}(c)] is squarelike, with sides approximately connected by $\textbf{Q}_{\rm AFM}$, as indicated by the red arrow.
Since Ce 4$f$ electrons are treated as core states, the calculated Fermi surface arises from conduction electrons that induce a Ruderman-Kittel-Kasuya-Yosida interaction between Ce 4$f$ local moments, leading to AFM spin fluctuations at $\textbf{Q}_{\rm AFM}$. \red{Although} the strong $k_z$ dependence of the DFT Fermi surface seems at odds with the quasi-2D character of the AFM fluctuations observed in our experiments, \red{this discrepancy suggests the dominance of in-plane correlations, which is crucial in theoretical models predicting the singlet-triplet transition in \CRA{} \cite{schertenleib2021unusual, skurativska2021spin, nogaki2021topological, mockli2021superconductivity, cavanagh2022nonsymmorphic, nogaki2022even, cavanagh2023pair, yanase2022topological, suh2023superconductivity, szabo2024effects}.}
Alternatively, as \CRA{} is a heavy fermion metal with 4$f$ states that hybridize with the conduction bands, the spin fluctuations at $\textbf{Q}_{\rm AFM}$ may arise from particle-hole excitations of the heavy electron bands.
\red{The quasi-2D AFM fluctuations observed in our neutron scattering measurements uniquely identifies $(\pi,\pi)$ as the primary nesting vector, among other plausible nesting vectors suggested by recent ARPES studies \cite{chen2024coexistence, wu2024fermi,chen2024exploring}.}

Resistivity and specific heat measurements reveal non-Fermi-liquid behavior at low $T$, which indicates proximity to a QCP \cite{khim2021field, hafner2022possible}. Our observations of anisotropic AFM fluctuations in \CRA{} suggest that this QCP has a nematic 2D AFM character, where the proximate ordered phase likely breaks the fourfold rotational symmetry of the underlying crystal lattice. Quantum critical fluctuations near a QCP are natural candidates as bosonic modes that drive superconductivity. Accordingly, our findings highlight the prominent role of AFM spin fluctuations in the superconductivity of \CRA, as in many other heavy-fermion superconductors. The momentum and energy dependence of the spin fluctuations in \CRA{} revealed in our work provide stringent constraints on models of its \orange{zero-field superconductivity}.

Time reversal and inversion symmetry in a metal enable even parity spin-singlet superconductivity and odd parity spin-triplet superconductivity, respectively. However, in superconductors lacking global inversion symmetry, electronic antisymmetric SOC, such as Rashba SOC, lift the spin degeneracy and mix spin-singlet and spin-triplet states \cite{nica2022multiple, gor2001superconducting}.
The unique local inversion symmetry breaking of \CRA{} naturally connects it to the previous proposal of noncentrosymmetric superconductivity in CeCoIn$_5$ multilayers \cite{yoshida2012pair}. 
Alternating Rashba SOC has been known to support distinct SC phases with spin-singlet and spin-triplet symmetries in the field-temperature phase diagram \cite{nica2022multiple, fischer2023superconductivity}. 
Experimental results in \CRA{} appear consistent with this picture, suggesting that the SC1 and high-field SC2 phases correspond to spin-singlet and spin-triplet pairing states, respectively.
However, the recent NMR study \cite{ogata2023parity} reported $T$-dependent Knight shifts, suggesting spin-singlet pairing in both SC phases, adding complexity to the Rashba SOC scenario. 
Since AFM spin fluctuations are typically seen in spin-singlet unconventional superconductors and ferromagnetic spin fluctuations are expected for spin-triplet superconductors,
our observation of AFM spin fluctuations at zero field is consistent with spin-singlet pairing in the SC1 phase. The fact that these AFM fluctuations extend to 1.2~meV $\gg k_{\rm B}T_{\rm C}$ suggests they will persist in the field-induced SC2 phase. In such a scenario, the SC2 phase should exhibit a prominent spin-singlet pairing component, consistent with the NMR measurement \cite{ogata2023parity}.
However, as research on UTe$_2$ revealed, AFM spin fluctuations can mediate spin-triplet superconductivity \cite{duan2021resonance}. \orange{Thus, additional theoretical and experimental efforts are necessary to elucidate the symmetry and mechanism of the SC2 phase. Nevertheless, our observation of AFM spin fluctuations places strong constraints on models of its field-induced superconductor-superconductor transition.}

We have investigated the magnetic excitations in \CRA{} using neutron scattering, revealing nematic quasi-2D AFM spin fluctuations centered at $\textbf{Q}_{\rm AFM}$ and extending up to 1.2 meV at low $T$. 
\red{Our neutron scattering measurements and DFT calculations indicate $\textbf{Q}_{\rm AFM}$ fluctuations result from the nesting of heavy quasi-2D Ce 4$f$ electrons,} which are also responsible for the heavy-fermion superconductivity in \CRA{}. These findings indicate a crucial role of AFM spin fluctuations in the superconductivity of \CRA{}, although the connection to the field-induced superconductor-superconductor transition remains to be understood.

{\it Acknowledgments} - We gratefully acknowledge the support from T. Xie, and R. Kumar during the experiments and the valuable discussions with S.-H. Baek, P. Dai, Y. Liu, E. Hassinger, and D. Agterberg. Work at the Institute for Quantum Matter, was supported by DOE, Office of Science, Basic Energy Sciences under Award No. DE-SC0019331 and DE-SC0024469. Research at Washington University was supported by the National Science Foundation (NSF) Division of Materials Research Award DMR-2236528. \red{Research at the University of Maryland was supported by the U.S. Department of Energy (DOE) Award No. DE-SC-0019154 (experimental preparation). J.P. acknowledges support from the Gordon and Betty Moore Foundation’s EPiQS Initiative through Grant No. GBMF4419.} C.B. was supported by the Gordon and Betty Moore Foundation EPIQS program under GBMF9456 and acknowledges the hospitality of the Instituto de Física at UNAM, México. Research at Zhejiang University was supported by the National Key R\&D Program of China (No. 2022YFA1402200). A portion of this research used resources at the Spallation Neutron Source, a DOE Office of Science User Facility operated by ORNL.

\bibliography{CRA}

\end{document}


\preprint{APS/123-QED}

\title{Supplemental Material for Antiferromagnetic Spin Fluctuations in the Spin-Triplet Superconductor Candidate \CRA}

\author{Tong Chen}
\email{tchen115@jhu.edu}

\author{Hasan Siddiquee}

\author{Qiaozhi Xu}

\author{Zack Rehfuss}

\author{Shiyuan Gao}

\author{Chris Lygouras}

\author{Jack Drouin}

\author{Vincent Morano}

\author{Keenan E. Avers}

\author{Christopher J. Schmitt}

\author{Andrey Podlesnyak}

\author{\red{Johnpierre Paglione}}

\author{Sheng Ran}
\email{rans@wustl.edu}

\author{Yu Song}
\email{yusong_phys@zju.edu.cn}

\author{Collin Broholm}
\email{broholm@jhu.edu}

\date{\today}

\maketitle


\section{\label{Supplementary : Sample alignment, and X-ray Laue Pattern}Sample alignment, and X-ray Laue Pattern}

\begin{figure}[h]
    \centering
    \includegraphics[width=0.9\textwidth]{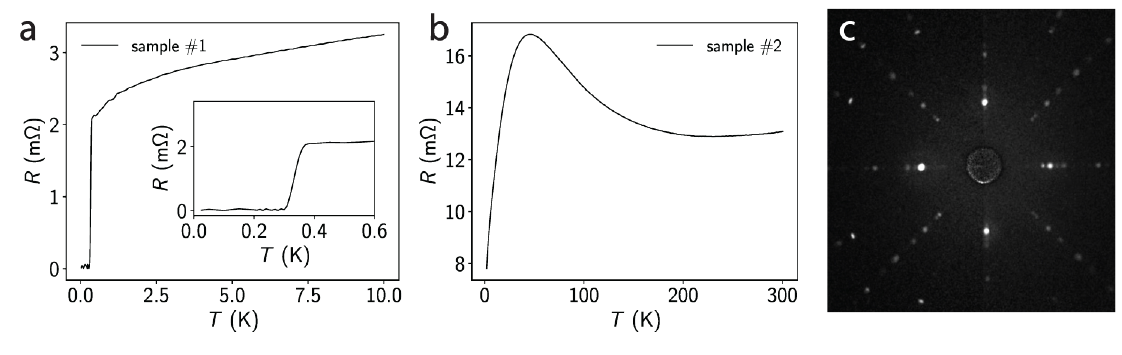}
    \caption{
    (a) Temperature-dependent resistivity of \CRA{}. The inset highlights the approximate 0.3~K superconductivity offset.
    (b) Temperature-dependent resistivity of \CRA{} up to 300~K.
    (c) X-ray Laue pattern of a \CRA{} single crystal, displaying the crystallographic orientation within the ab-plane.
    } \label{fig:sfig1}
\end{figure}

\begin{figure}[h]
    \centering
    \includegraphics[width=0.9\textwidth]{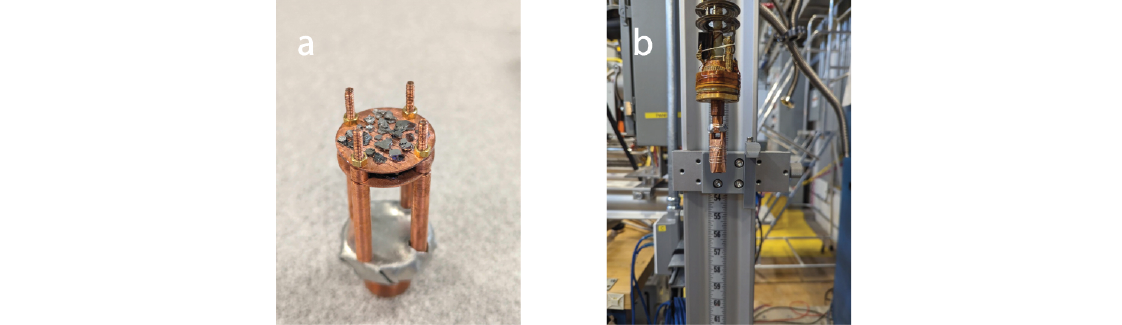}
    \caption{
    (a) Photo depicting the assembly aligned in the $(hk0)$ scattering plane for CNCS.
    (b) Photo showing the assembly with an oxygen-free high conductivity (OFHC) copper shielding attached to the cold finger of a dilution refrigerator.
    } \label{fig:sfig2}
\end{figure}

High-quality single crystals of \CRA{} were synthesized using Bi flux methods, as described in \cite{siddiquee2023pressure}. Representative resistivity plots are shown in Figures~\ref{fig:sfig1}(a,b). These single crystals typically manifest as octagonal plates with dimensions of 1.0~mm $\times$ 1.0~mm $\times$ 0.2~mm. Natural edges align along either (110) or (100) directions, while the out-of-plane direction corresponds to (001), the $\textbf{c}$-axis (Fig.~\ref{fig:sfig1}(c)). A total mass of 0.9~g, comprising approximately 120 pieces of \CRA{} single crystals, was co-aligned on two oxygen-free high conductivity (OFHC) copper plates using CYTOP for attachment. Thermally shielded by OFHC copper foil secured to the OFHC base, the sample mount was attached to the cold finger of a dilution refrigerator with 80~$\mu$W cooling power at 100~mK (Fig.~\ref{fig:sfig2}).

\section{\label{appendix: Absolute Normalization} Absolute Normalization}

The monitor normalized scattering intensity obtained from initial data reduction is related to the scattering cross-section as follows: 
\begin{equation}
{\cal I}({\bf Q},\omega)=N{\cal C}\frac{k_i}{k_f}\overline{\frac{\d^2\sigma}{\d\Omega \d E_f}}({\bf Q},\omega)
\end{equation}
Here $N{\cal C}$ is a configuration-dependent normalization constant that is proportional to the number of unit cells in the sample ($N$). The bar indicates convolution with the instrumental resolution function: 
\begin{equation}
\overline{\frac{\d^2\sigma}{\d\Omega \d E_f}}({\bf Q},\omega)=\int \d^3{\bf Q}^\prime \hbar \d\omega^\prime 
{\cal R}_{{\bf Q},\omega}({\bf Q-Q^\prime},\omega-\omega^\prime)\frac{\d^2\sigma}{\d\Omega \d E_f}({\bf Q^\prime},\omega^\prime)
\end{equation}
${\cal R}_{{\bf Q},\omega}$ is sharply peaked when its arguments vanish and unity normalized: 
\begin{equation}
\int \d^3{\bf Q^\prime}\hbar \d\omega^\prime {\cal R}_{{\bf Q},\omega}({\bf Q-Q^\prime},\omega^\prime)\equiv 1
\end{equation}

Absolute normalization of the scattering cross-section was achieved using the integrated intensity of the (110) nuclear Bragg reflection. This has the advantage of being unaffected by sample absorption for the elastic channel. The Bragg scattering cross section per unit cell of volume $v_N$ is given by (Fig.~\ref{fig:sfig3}(a)):
\begin{equation}
\begin{aligned}
\left(\frac{\d^2\sigma}{\d \Omega \d E_f}\right)_{\rm el}^{\rm nuc}=\frac{(2\pi)^3}{v_N} \sum_{\bm{\tau}} |\mathbb{F}_N(\bm{\tau})|^2\delta({\bf Q}-\bm{\tau})\delta(\hbar\omega),
\label{Eq:nuc}
\end{aligned}
\end{equation}
where $\mathbb{F}_N(\bm{\tau})$ is the structure factor given by:
\begin{equation}
\begin{aligned}
\mathbb{F}_N(\bm{\tau})=\sum_{\bf d} b_{\bf d} \textup{exp}(-2W_{\bf d}({\bf Q})) \textup{exp}(i\bm{\tau}\cdot{\bf d}).
\end{aligned}
\end{equation}
Here $b_{\bf d}$ is the bound coherent scattering length and $\textup{exp}(-2W_{\bf d}({\bf Q}))$ is the Debye-Waller factor for the atom on the site $\bf d$ within the unit cell with volume $v_N$. The scattering intensity near the $\bm{\tau}=$(110) Bragg peak is thus given by
\begin{equation}
{\cal I}_N({\bf Q},\omega)=N{\cal C}\frac{(2\pi)^3}{v_N}  |\mathbb{F}_N(\bm{\tau})|^2 {\cal R}_{\bm{\tau},0}({\bf Q-\bm{\tau}},\omega).
\end{equation}
Using the normalization property of the resolution function we find
\begin{equation}
    N{\cal C}=\int_{{\bf Q\approx \bm{\tau}},\omega\approx 0} \d^3{\bf Q} \hbar \d\omega {\cal I}_N({\bf Q},\omega)\left(\frac{(2\pi)^3}{v_N} |\mathbb{F}_N(\bm{\tau})|^{2}\right)^{-1}
\end{equation}

We thus obtain the following expression relating the measured neutron scattering intensity to the resolution smeared absolute normalized scattering cross-section per unit cell: 
\begin{equation}
    \frac{k_i}{k_f}\overline{\frac{\d^2\sigma}{\d\Omega \d E_f}}({\bf Q},\omega)=\frac{{\cal I}({\bf Q},\omega)}{\int_{\bm{\tau}}  \d^3{\bf Q} \hbar \d\omega {\cal I}_N({\bf Q},\omega)}\frac{(2\pi)^3}{v_N} |\mathbb{F}_N(\bm{\tau})|^2
\end{equation}

If we denote by $\overline{{\cal I}_N}(110)$ the average (110) nuclear Bragg intensity throughout a volume of ${\bf Q}-\omega$ space that exceeds the resolution volume and has dimensions $\Delta h, \Delta k, \Delta l$ in reciprocal lattice units and $\Delta \hbar\omega$ in energy, this expression simplifies to 
\begin{equation}
    \frac{k_i}{k_f}\overline{\frac{\d^2\sigma}{\d\Omega \d E_f}}({\bf Q},\omega)=\frac{{\cal I}({\bf Q},\omega)}{\overline{{\cal I}_N}(110)\Delta h\Delta k\Delta l \Delta \hbar\omega}|\mathbb{F}_N(\bm{\tau})|^2
\label{Eq:Inorm}
\end{equation}
for orthogonal reciprocal lattice vectors.


\section{\label{Supplementary : Absorption Correction}Absorption Correction}

When a neutron beam travels through the sample, the decrease in scattering intensity due to the absorption cross-section is quantified by the transmission coefficient:
\begin{equation}
\begin{aligned}
A = \frac{1}{V} \int {\rm exp}(-\mu T) {\rm d} V,
\end{aligned}
\end{equation}
where $\mu$ is the attenuation length, and the integration is over the volume of the crystal. Here, $T$ denotes the path length of the neutron beam within the crystal, encompassing the cumulative path lengths of both the incident and outgoing neutrons \cite{hahn1983international, arnold2014mantid}.

For the scattered neutron intensity within the equatorial plane of a cylinder with radius $R$ in the neutron beam, the expression for the transmission coefficient is:
\begin{equation}
\begin{aligned}
A(\sigma_r,\rho,\lambda_i,\lambda_f,R,\theta) = \frac{1}{\pi R^2} \int_0^R \int_0^{2\pi} {\rm exp}(&-\mu(\lambda_i)\{[R^2-r^2{\rm sin}^2(\phi+\theta)]^{1/2}+r{\rm cos}(\phi+\theta)\}\\
&-\mu(\lambda_f)\{[R^2-r^2{\rm sin}^2(\phi-\theta)]^{1/2}-r{\rm cos}(\phi-\theta)\})r{\rm d}r{\rm d}\phi,\\
\mu(\lambda) = \rho * \sigma(\lambda) = &\rho * \sigma_r \frac{\lambda}{1.7982\ {\rm \AA}},
\label{eq:trans_coff}
\end{aligned}
\end{equation}
where $\rho$ is the atomic number density of the sample and $\sigma_r$ is the absorption cross-section for the reference wavelength of 1.7982\ {\rm \AA}. Thus, the transmission coefficient $A$ is a function of $\sigma_r$,  $\rho$, the wavelengths of the incident and outgoing neutrons $\lambda_i$ and $\lambda_f$, the radius of the cylinder sample $R$, and the scattering angle $\theta$. The calculated transmission coefficients agree with those obtained through numerical integration \cite{dwiggins1975rapid}. For \CRA,  
\begin{equation}
\begin{aligned}
\rho=\frac{M}{V} \frac{N_{\rm A}}{m} = 9.05~{\rm g/cm}^3 * \frac{6.023 \times 10^{23}~{\rm mol}^{-1}}{492.109~{\rm g/mol}} = 0.01108~{\rm \AA}^{-3},
\end{aligned}
\end{equation}
where $\frac{M}{V}$ is the density, $N_{\rm A}$ is Avogadro’s constant, and $m$ is the molecular weight. From the data on the NIST website \cite{NISTwebsite, sears1992neutron}, we have $\sigma_r=299.23$~barn/f.u. and in our experimental setup, the sample assembly resembles a cylinder with $R = 3.5$~mm. With $\sigma_r=299.23$~barn/f.u., $\rho=0.01108~{\rm \AA}^{-3}$, $\lambda_i=4.964$~\AA, and $R = 3.5$~mm, the transmission coefficient $A$ (Eq.~\ref{eq:trans_coff}) depends on $\theta$ and $\lambda_f$, which is thus a function of $\textbf{Q}$ and $\omega$. 

Since we performed the absolute normalization using the (110) nuclear Bragg peak, the absorption correction is conducted by comparing the absorption coefficient  $A(\textbf{Q},\omega)$ with that of the (110) peak. After the absorption correction, Equation~\ref{Eq:Inorm} becomes:
\begin{equation}
\begin{aligned}
    \frac{k_i}{k_f}\overline{\frac{\d^2\sigma}{\d\Omega \d E_f}}({\bf Q},\omega)=\frac{{\cal I}({\bf Q},\omega)}{\overline{{\cal I}_N}(110)\Delta h\Delta k\Delta l \Delta \hbar\omega} \frac{A(\bm{\tau}, 0)}{A(\textbf{Q},\omega)}|\mathbb{F}_N(\bm{\tau})|^2,\\
\end{aligned}
\end{equation}
where $A(\bm{\tau}, 0) = A(\bm{\tau}=(110), \omega=0)= 0.0671$.


\section{\label{appendix: Upper Bound of Ordered Moment}Upper Bound of Ordered Moment}

While our experiment did not detect elastic magnetic Bragg diffraction in \CRA{}, the small sample neutron absorbing sample means that this does not set a particularly strict limit on the staggered magnetization of possible magnetic ordering in the sample. To determine the limits on any ordered moment set by our experiment we start by the following general expression for the elastic magnetic scattering cross section. 
\begin{equation}
\begin{aligned}
\left(\frac{\textup{d}^2 \sigma}{\textup{d} \Omega \textup{d} E_f}\right)_{\rm el}^{\rm mag}
&=
(\frac{\gamma r_0}{2})^2 |gF({\bf Q})|^2 \sum_{\alpha\beta} (\delta_{\alpha\beta} - \hat{Q}_\alpha\hat{Q}_\beta) 
\frac{1}{N}\sum_{RdR'd'} \langle S_{Rd}^\alpha \rangle \langle S_{R'd'}^\beta \rangle e^{i{\bf Q}\cdot(\bm{R}+\bm{d}-\bm{R}'-\bm{d}')} \delta(\hbar\omega)\\
&= 
(\frac{\gamma r_0}{2})^2 |gF({\bf Q})|^2 (1-\hat{Q}_z^2) 
\frac{1}{N}\sum_{R d R'd'}
\langle S_{R d}^z \rangle
\langle S_{R' d'}^z \rangle
e^{i{\bf Q} \cdot(\bm{R}-\bm{R}'+\bm{d}-\bm{d}')}\delta(\hbar\omega)
\end{aligned}
\end{equation}
Here $\gamma=1.913$ is the ratio between the neutron dipole moment and the nuclear Bohr magneton, $r_0=2.8179$~fm is the classical electron radius, $|F({\bf Q})|$ is the cerium form factor, $\hat{Q}_z$ is the $z$-component of the unit scattering vector $\bm{\hat{Q}}$. The factor $1/N$ where $N$ is the number of unit cells of volume $v_N$ ensures consistent normalization with Eq.~\ref{Eq:nuc}. For simplicity, we assume that any magnetic order is uniaxial and polarized along the $\bf c$ axis. Other assumptions will modify the inferred limits by factors of order unity. 

\begin{figure}[h]
    \centering
    \includegraphics[width=0.9\textwidth]{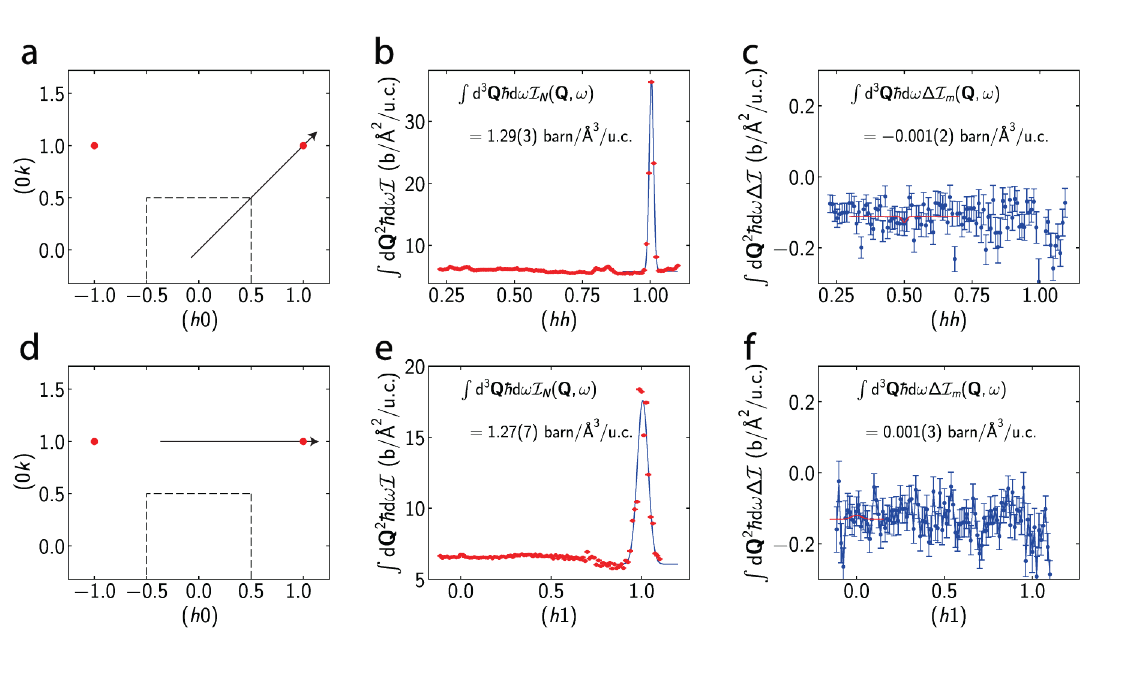}
    \caption{
    (a) Schematic of the $(hk)$ plane and the cut along the $(hh)$ direction. 
    (b) Elastic scattering along $(hh)$. Since no discernible difference was observed between 0.08~K and 0.8~K, data are averaged to enhance statistics. 
    (c) Temperature difference (0.08~K - 0.8~K) of the elastic scattering along $(hh)$. 
    Data in (b,c) are averaged along $(k\bar{k})$ and $l$ in windows of $\pm 0.10$ and $\pm0.5$~r.l.u., respectively. The energy window is from -0.2~meV to 0.2~meV. Gaussian peaks are fitting for the (110) nuclear peak (b) and putative magnetic scattering at $(\frac{1}{2}\frac{1}{2})$ (c).
    (d) Schematic of the $(hk)$ plane and the cut along the $(h1)$ direction. 
    (e) Elastic scattering along $(h1)$. 0.08~K and 0.8~K data are averaged. 
    (f) Temperature difference (0.08~K - 0.8~K) of the elastic scattering along $(h1)$. 
    Data in (e,f) are averaged along $(0k)$ and $l$ in windows of $\pm 0.15$ and $\pm0.5$~r.l.u., respectively. The energy window is from -0.2~meV to 0.2~meV. Gaussian peaks are fitted for the (110) nuclear peak (e) and putative magnetic scattering at $(01)$ (f).
    } \label{fig:sfig3}
\end{figure}

For three-dimensional magnetic orders, we obtain
\begin{equation}
\begin{aligned}
\left(\frac{\textup{d}^2 \sigma}{\textup{d} \Omega \textup{d} E_f}\right)_{\rm el}^{\rm mag}&= 
(\frac{\gamma r_0}{2})^2 |F({\bf Q})|^2 (1-\hat{Q}_z^2) 
(m_z/\mu_B)^2 
\frac{(2\pi)^3}{v_m} \frac{N_m}{N}\sum_{\tau_{m}} |\mathbb{F}_m(\bm{\tau}_{m})|^2 \delta^3({\bf Q} - \bm{\tau}_{m})\delta(\hbar\omega )\\
\end{aligned}
\end{equation}
Here $N_m$ is the number of magnetic unit cells of volume $ v_m$, $\bm{\tau}_{m}$ are reciprocal lattice vectors of the magnetic unit cell, $m_z$ is the ordered magnetic dipole moment, and $\mathbb{F}_m({\bf Q})$ is the scalar magnetic structure factor:
\begin{equation}
\mathbb{F}_m({\bf Q})=\sum_{{\bf d}_m}\textup{exp}(-i{\bf Q}\cdot {\bf d}_m).
\end{equation}
Here ${\bf d}_m$ indicates position vectors for magnetic ions within the magnetic unit cell and we have suppressed the Debye-Waller factor for simplicity. From the analogous forms of the magnetic and nuclear Bragg diffraction cross-section we readily obtain the following: 
\begin{equation}
    m_z^2=\frac{\int_{\bm{\tau}_m}  \d^3{\bf Q} \hbar \d\omega {\cal I}_m({\bf Q},\omega)}{\int_{\bm{\tau}}  \d^3{\bf Q} \hbar \d\omega {\cal I}_N({\bf Q},\omega)}\cdot
    \frac{A(\bm{\tau},0)}{A(\bm{\tau_m},0)} \cdot
    \left(\frac{v_m}{v_N}\right)^2 \cdot
    \frac{|\mathbb{F}_N(\bm{\tau})|^2}{(\frac{\gamma r_0}{2})^2|F(\bm{\tau}_m)|^2(1-\hat{\tau}_{mz}^2)|\mathbb{F}_m({\bm{\tau}_m})|^2} \mu_B^2.
    \label{3Dmomentsize}
\end{equation}
Here we have used $N_mv_m=Nv_N$ and introduced the sample absorption correction. In Figure~\ref{fig:sfig3}, we show the elastic scattering along the $(hh)$ and $(h1)$ directions, from which we calculate $\int_{\bm{\tau}_m}  \d^3{\bf Q}\hbar \d\omega {\cal I}_m({\bf Q},\omega)$ for putative magnetic orders at $(\frac{1}{2}\frac{1}{2})$ and $(01)$, respectively.

For quasi-two-dimensional magnetic order, there are no magnetic correlations between cerium planes and the diffraction cross-section takes this form
\begin{equation}
\begin{aligned}
\left(\frac{\textup{d}^2 \sigma}{\textup{d} \Omega \textup{d} E_f}\right)_{\rm el}^{\rm mag}&= 
(\frac{\gamma r_0}{2})^2 |F({\bf Q})|^2 (1-\hat{Q}_z^2) 
(m_z/\mu_B)^2  
\frac{(2\pi)^2}{A_m} \frac{N_m}{N}\sum_{\tau_{m}} |\mathbb{F}_m(\bm{\tau}_{\parallel m})|^2 \delta^2({\bf Q}_\parallel - \bm{\tau}_{\parallel m}) \delta (\hbar\omega)\\
\label{quasi2D}
\end{aligned}
\end{equation}
Here $N_m$ is the number of 2D magnetic unit cells with area $A_m$ and $\mathbb{F}_m$ is the corresponding scalar magnetic structure factor. Note that $\frac{1}{2}N_m A_m c=Nv_n$ where the factor of a half arises because there are two cerium layers per nuclear unit cell.  

Since the \orange{CNCS} instrument used in this work only covers a relatively small range of wave vector transfer perpendicular to the scattering plane, we can approximate the resolution function with a factorized form: 
\begin{equation}
    {\cal R}_{{\bf Q},\omega}(\delta{\bf Q},\delta \omega)\approx {\cal R}^\parallel_{{\bf Q}_\parallel,\omega}(\delta{\bf Q}_\parallel,\delta \omega) {\cal R}^\perp(\delta{\bf Q}_z),
\end{equation}
where both ${\cal R}^\parallel$ and ${\cal R}^\perp$ are unity normalized over their domain. Convoluting Eq.~\ref{quasi2D} with this form of the resolution function, we obtain the following approximate form for the resolution smeared diffraction cross-section: 
\begin{equation}
\begin{aligned}
\left(\overline{\frac{\textup{d}^2 \sigma}{\textup{d} \Omega\textup{d} E_f}}\right)_{\rm el}^{\rm mag}&= 
(\frac{\gamma r_0}{2})^2 |F({\bf Q})|^2 (1-\hat{Q}_z^2) 
(m_z/\mu_B)^2 
\frac{(2\pi)^2}{A_m} \frac{N_m}{N}|\mathbb{F}_m(\bm{\tau}_{\parallel m})|^2 {\cal R}^\parallel_{{\bf Q}_\parallel,\omega}({\bf Q}_\parallel-\bm{\tau}_{\parallel m},\omega)\\
\end{aligned}
\end{equation}

This finally leads to an expression for the staggered magnetization: 
\begin{equation}
    m_z^2=\frac{\int_{\bm{\tau}_m}  \d^3{\bf Q}\hbar \d\omega {\cal I}_m({\bf Q},\omega)(|F({\bf Q})|^2(1-\hat{Q}_z^2))^{-1}}{\int_{\bm{\tau}}  \d^3{\bf Q} \hbar \d\omega {\cal I}_N({\bf Q},\omega)}\cdot 
    \frac{A(\bm{\tau},0)}{A(\bm{\tau_m},0)}  \cdot
    \left(\frac{A_m c}{v_N}  \right)^2 \cdot
    \frac{ (c^*/\Delta Q_z)  |\mathbb{F}_N(\bm{\tau})|^2}{(\frac{\gamma r_0}{2})^2 2|\mathbb{F}_m({\bm{\tau}_{\parallel m}})|^2}\mu_B^2
    \label{2Dmomentsize}
\end{equation}
Here $\Delta Q_z$ is the range of integration along the $\bf c$ direction for the $\bf Q$ integration in the numerator and we have used $N_m/N=2(v_N/A_mc)$. Table~\ref{table:s1} shows \orange{the in-plane area of the magnetic unit cell $A_m$}, \orange{the c-axis lattice parameter of the magnetic unit cell} $c_m$, polarization factor $(1-\hat{Q}_z^2)$, magnetic form factor $|F(\textbf{Q})|$, $|\mathbb{F}_m|$, and corresponding $m_z^2$ and upper bounds for putative magnetic orders at specific wavevectors. Note that there are magnetic peaks at both $l=-0.5$ and $0.5$ for $(\frac{1}{2}\frac{1}{2}\frac{1}{2})$ and $(01\frac{1}{2})$ order, but this factor is coincidentally compensated by the fact that the $l$ coverage is -0.5 to 0.5~r.l.u. 

\begin{table}[h]
\begin{center}
\caption{Magnetic ordering and upper limit of moment determined from the corresponding analysis of our diffraction data.}
\label{table:s1}
\begin{tabular}{|c|c|c|c|c|c|c|c|c|} 
\hline
wavevector ${\bf Q}$ &$A_N/A(\textbf{Q},\omega)$ & $A_m$ & $c_m$ & $(1-\hat{Q}_z^2)$ & $|F(\textbf{Q})|$ & $|\mathbb{F}_m|^2$ & $m_z^2$($\mu_{\rm B}^2$) & $m_{\parallel,max}$ ($\mu_{\rm B}$)\\
\hline
($\frac{1}{2}\frac{1}{2}0$) & 2.46 & 4a & c & 1 & 0.937 & 64 & 0.00(1) & 0.10\\
($\frac{1}{2}\frac{1}{2}\frac{1}{2}$)& 2.46 & 4a & 2c & 0.914 & 0.932 & 128 & -0.01(3) & 0.14\\
(010) & 1.65 & a & c & 1 & 0.880 & 4 & 0.01(1) & 0.14\\
(01$\frac{1}{2}$) & 1.65 & a & 2c & 0.955 & 0.875 & 8 & 0.01(3) & 0.22\\
quasi-2D ($\frac{1}{2}\frac{1}{2}$) & 2.46 & 4a & c & $> 0.914$ & $> 0.932$ & 16 & -0.01(3) & 0.14 \\
 \hline
\end{tabular}
\end{center}
\end{table}


\section{\label{Supplementary : CYTOP Scattering}Scattering from CYTOP}

\begin{figure}[h]
    \centering
    \includegraphics[width=0.9\textwidth]{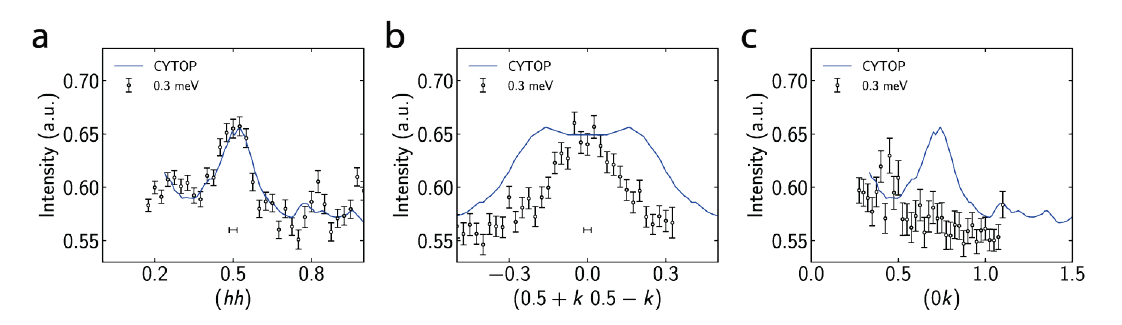}
    \caption{
    (a) Constant energy cuts along the longitudinal $(hh)$ direction at 0.3~meV. The averaging windows along $(k\bar{k})$ and $l$ directions are $\pm 0.15$, and $\pm 0.5$ r.l.u., respectively. 
    (b) Constant energy cuts along the transverse $(k\bar{k})$ direction at 0.3~meV. The averaging windows along $(hh)$ and $l$ directions are $\pm 0.1$, and $\pm 0.5$ r.l.u., respectively.
    (c) Constant energy cuts along the longitudinal $k$ direction at 0.3~meV. The averaging windows along $h$ and $l$ directions are $\pm 0.2$, and $\pm 0.5$ r.l.u., respectively.
    The energy window is $\pm 0.1$ meV.
    The horizontal error bars represent the resolution width.
    } \label{fig:sfig4}
\end{figure}

We used the fluoropolymer-based adhesive CYTOP to attach the crystals. CYTOP has a powder-like diffraction peak around 1.1~$\textup{\AA}$, which practically coincides with ${\bf Q}_{\rm AF} = (\frac{1}{2}\frac{1}{2})$ in \CRA, as shown in Figure~\ref{fig:sfig4}(a). Therefore, it is crucial to examine whether the peak at ${\bf Q}_{\rm AF}$ could be from CYTOP. We can rule this out because scattering from CYTOP is approximately isotropic and independent of sample rotation while magnetic scattering is four-fold rotational symmetric. 

To compare the scattering from CYTOP with the excitation spectrum, we scaled the $(hh)$ cut of the CYTOP scattering to that of the peak at 0.3~meV and constructed the $(k\bar{k})$ and $(0k)$ cuts for CYTOP (Fig.~\ref{fig:sfig4}(b,c)). The CYTOP scattering is wider than the observed excitations centered at $\textbf{Q}_{\rm AF} = (\frac{1}{2}\frac{1}{2})$ in the $(k\bar{k})$ cut. In addition, CYTOP also shows a peak in the $(0k)$ cut, which is absent in the measured spectrum.

\section{\label{Supplementary : Nesting Function} Nesting Function}

\begin{figure}[h]
    \centering
    \includegraphics[width=0.8\textwidth]{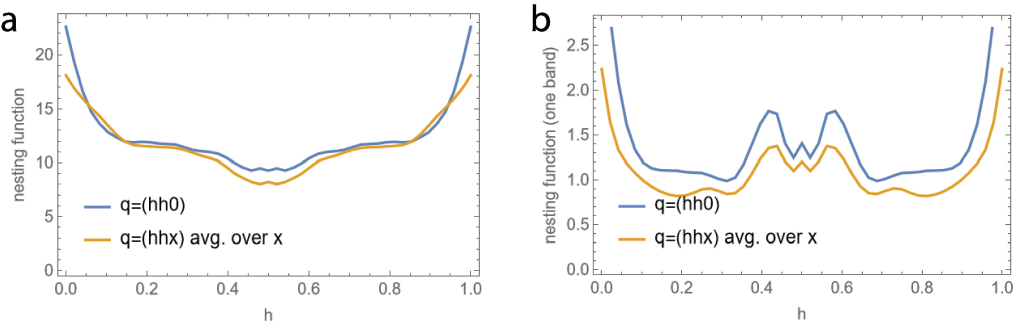}
    \caption{
    (a) Wavevector dependence of nesting function from DFT calculations. 
    (b) Wavevector dependence of nesting function for one band. Blue curves are calculated at $l=0$ and orange curves are calculated with averaged $l$.
    } \label{fig:sfig5}
\end{figure}

\red{The nesting function was obtained from DFT calculations, yet it does not display a pronounced peak at ($\frac{1}{2}\frac{1}{2}$) (Fig.~\ref{fig:sfig5}(a)). This absence of a clear nesting feature is common in heavy-fermion superconductors. Such deviations from expected nesting peaks underscore the complex interplay of electronic structure and interactions in these materials, which may not be fully captured by standard DFT methods alone.}

%
